\documentclass[12pt,a4paper]{article}
\usepackage{latexsym,graphicx,verbatim,amssymb,amsfonts}

\catcode`\@=11
\def\marginnote#1{}
\newcount\hour
\newcount\minute
\newtoks\amorpm
\hour=\time\divide\hour by60
\minute=\time{\multiply\hour by60 \global\advance\minute by-\hour}
\edef\standardtime{{\ifnum\hour<12 \global\amorpm={am}%
        \else\global\amorpm={pm}\advance\hour by-12 \fi
        \ifnum\hour=0 \hour=12 \fi
        \number\hour:\ifnum\minute<10 0\fi\number\minute\the\amorpm}}
\edef\militarytime{\number\hour:\ifnum\minute<10 0\fi\number\minute}
\def\draftlabel#1{{\@bsphack\if@filesw {\let\thepage\relax
   \xdef\@gtempa{\write\@auxout{\string
      \newlabel{#1}{{\@currentlabel}{\thepage}}}}}\@gtempa
   \if@nobreak \ifvmode\nobreak\fi\fi\fi\@esphack}
        \gdef\@eqnlabel{#1}}
\def\@eqnlabel{}
\def\@vacuum{}
\def\draftmarginnote#1{\marginpar{\raggedright\scriptsize\tt#1}}
\def\draft{\oddsidemargin -.5truein
        \def\@oddfoot{\sl preliminary draft \hfil
        \rm\thepage\hfil\sl\today\quad\militarytime}
        \let\@evenfoot\@oddfoot \overfullrule 3pt
        \let\label=\draftlabel
        \let\marginnote=\draftmarginnote
   \def\@eqnnum{(\theequation)\rlap{\kern\marginparsep\tt\@eqnlabel}%
\global\let\@eqnlabel\@vacuum}  }

\def\preprint{\twocolumn\sloppy\flushbottom\parindent 1em
        \leftmargini 2em\leftmarginv .5em\leftmarginvi .5em
        \oddsidemargin -.5in    \evensidemargin -.5in
        \columnsep 15mm \footheight 0pt
        \textwidth 250mmin      \topmargin  -.4in
        \headheight 12pt \topskip .4in
        \textheight 175mm
        \footskip 0pt
        \def\@oddhead{\thepage\hfil\addtocounter{page}{1}\thepage}
        \let\@evenhead\@oddhead \def\@oddfoot{} \def\@evenfoot{} }

\def\titlepage{\@restonecolfalse\if@twocolumn\@restonecoltrue\onecolumn
     \else \newpage \fi \thispagestyle{empty}\c@page\z@ 
        \def\thefootnote{\fnsymbol{footnote}} }

\def\endtitlepage{\if@restonecol\twocolumn \else  \fi
        \def\thefootnote{\arabic{footnote}}
        \setcounter{footnote}{0}}  

\catcode`@=12
\relax

\def\bea{\begin{array}}
\def\bem{\begin{displaymath}}
\def\beq{\begin{equation}}

\def\eea{\end{array}}
\def\eem{\end{displaymath}}
\def\eeq{\end{equation}}

\def\Im{\mathop{\rm Im}}

\def\ov{\overline}

\def\Re{\mathop{\rm Re}}

\def\s2w{\sin^2 \theta_W}

\relax
\def\dalpha{{\dot\alpha}}

\def\crbig{\\\noalign{\vspace {3mm}}}

\relax

\begin{document}
\topmargin-.7cm
%
%
%
%
\begin{titlepage}

\vspace{1.9cm}

\begin{center}{\Large\bf
On supergravity theories, after \boldmath{$\sim40$} years  }
\vspace{1.8cm}

{\large\bf Jean-Pierre Derendinger}

\vspace{1.3cm}
Albert Einstein Center for Fundamental Physics, \\
Institute for Theoretical Physics, Bern University \\
Sidlerstrasse 5, CH--3012 Bern, Switzerland \\
{\small\sl derendinger@itp.unibe.ch}
\end{center}
\vspace{1.9cm}

\begin{center}
{\large\bf Abstract}
\end{center}
\begin{quote}
An introduction to and a partial review of supergravity theories is given, insisting on 
concepts and on some important technical aspects. 
Topics covered include elements of global supersymmetry, a derivation of the simplest ${\cal N}=1$
supergravity theory, a discussion of ${\cal N}=1$ matter--supergravity couplings, of the scalar sector and 
of some simple models. Space-time is four-dimensional. 
\end{quote}

\vspace{1.8cm}

\begin{center}
{\it
Presented at DISCRETE 2014, the Fourth Symposium on Prospects \\ in the Physics of Discrete Symmetries, \\
King's College, London, December 2014. \\
Published in the Proceedings.}
\end{center}

\end{titlepage}
\renewcommand{\theequation}{\thesection.\arabic{equation}}
\setcounter{footnote}{0}
\setcounter{page}{0}
\setlength{\baselineskip}{.6cm}
\setlength{\parskip}{.2cm}
\newpage
%
%

\section{Foreword}

Almost forty years ago, in 1974 and soon after, most of the important properties of quantum field 
theories with linear supersymmetry\footnote{In four dimensions.} were displayed in a brilliant series of papers. 
These fundamental results include what is now called the Wess-Zumino model \cite{WZ}, 
super-Yang-Mills theory \cite{FZ1}\footnote{And later on \cite{BSS}.}, exceptional
renormalization properties \cite{IZ}, spontaneous supersymmetry breaking \cite{FI, O'R}, 
the current structure \cite{FZ2} and the development of superspace and superfield techniques \cite{SS, FWZ}.

Two attractive complementary aspects of supersymmetry were recognized. Firstly, it is an extension compatible 
with quantum mechanics of the Poincar\'e algebra, the symmetry of relativistic field theory \cite{GL, HLS, IZ}. Secondly,
it can be implemented in the Standard Model of particle interactions and be experimentally tested \cite{F}, leading to 
numerous models and analysis and, now, to LHC (preliminary and future) results.

Almost forty years ago, in 1976, the first supergravity theory, the gauge theory of supersymmetry, was invented by
Ferrara, Freedman and van Nieuwenhuizen \cite{FFVN} and by Deser and Zumino
\cite{DZ}, opening decades of developments which installed supergravity at the meeting point of two independent approaches to the unification program. 

On the low energy side, the success of the Standard Model 
emphasizes the enigma of the large hierarchical ratio $M_P/M_W$. Global supersymmetry helps with his capacity to
forbid destabilizing quantum corrections \cite{IZ, GSR}. But it does not spontaneously break, as observations
obviously require, and this is where supergravity helps, by proposing a source for supersymmetry breaking
effects at low energies \cite{BFS, ACN, GG} and also in proposing a scheme to radiatively induce a small scale 
($M_W$) from supersymmetry breaking at a much higher scale \cite{IR, AGCW, AGPW, IL, EHNT}.

On the high energy scale, at present, the only concrete proposal for a coherent description of all particle interactions
including gravitation involves models based on strings with scale close or related to the 
Planck scale. Coherence and stability of these models require some amount of supersymmetry. 
Supergravity appears then as the natural tool to describe the light sector of superstring theories, at energy
scales where string excitations decouple and where supersymmetry breaking should be able to generate
the so-called soft breaking terms needed in the supersymmetric Standard Model \cite{BFS, ACN, GG}.

More pragmatically, the possible relevance of supersymmetry to particle physics, in relation with the weak 
interaction scale,
is concrete and testable at LHC (and some other) experiments. This is maybe a strong enough motivation.

Over the years, supergravity theories have of course found many other more theoretical roles, in gauge/gravity 
dualities, to understand the structure of gravitational scattering amplitudes, in the description 
of superstring solutions and compactifications, in studies of black holes, solutions of gravitational theories 
with tensor fields, attractors, 
in sigma models living on particular geometries \dots

This contribution tries to provide an introduction to (four-dimensional) supergravity and 
a detailed discussion of some aspects of simple theories and of their applications. It begins with basic 
aspects of global supersymmetry (section \ref{global}), with a detailed discussion of the role of auxiliary
fields. The simplest ${\cal N}=1$ supergravity and its Anti-de Sitter deformation are then derived (section \ref{local}).
After a brief discussion of theories with extended supersymmetry, general properties of the ${\cal N}=1$
supergravity--matter couplings are described (section \ref{matter}), with particular attention to the scalar 
potential and the gravitino sector. The examples of simple no-scale or dilaton supergravities are the subject
of section \ref{noscale}.

\section{Elements of global supersymmetry} \label{global}
\setcounter{equation}{0}

Relativistic quantum field theories and strong or electroweak interactions 
are invariant under global transformations of the Poincar\'e group, {\it i.e.} under Lorentz 
(proper and orthochronous) transformations and translations. In other words, there are Lorentz generators 
$M^{\mu\nu}=-M^{\nu\mu}$ and translation generators $P^\mu$ acting on coordinates and fields and leaving
the dynamical equations unchanged. Translation generators are universal, $P_\mu = -i\partial_\mu$. 
Lorentz generators act on coordinates according to
\beq
\delta x^\mu = {i\over2} \omega_{\rho\sigma}M^{\rho\sigma} x^\mu = - \omega^{\mu\nu}x_\nu, \qquad\qquad
M^{\rho\sigma} = -i( x^\rho\partial^\sigma - x^\sigma\partial^\rho).
\eeq
They act on fields with generators in representations depending on the spins of the fields. For a set of fields $\Phi(x)$,
\beq
\delta \Phi(x) = {i\over2} \omega_{\rho\sigma} \Sigma^{\rho\sigma}\Phi(x) - \delta x^\mu \partial_\mu\Phi(x)
= {i\over2} \omega_{\rho\sigma}M^{\rho\sigma}\Phi(x)
\eeq
and the information on spins is in the choice of linear operators $\Sigma^{\mu\nu}$.
These variations represent the Poincar\'e Lie algebra
\beq
\label{P4}
\begin{array}{rcl} 
[M^{\mu\nu},M^{\rho\sigma}]&=& -i\left( \eta^{\mu\rho}M^{\nu\sigma}+ 
\eta^{\nu\sigma}M^{\mu\rho}-\eta^{\mu\sigma}M^{\nu\rho} 
-\eta^{\nu\rho}M^{\mu\sigma} \right) ,  \crbig
[P^\mu,M^{\nu\rho}]&=& i\left( \eta^{\mu\nu}P^\rho-\eta^{\mu\rho}P^\nu\right), 
\crbig
[P^\mu,P^\nu]&=& 0. 
\end{array}
\eeq

\subsection{The Poincar\' e superalgebra}

Global supersymmetry is an extension compatible with quantum field theory of Poin\-ca\-r\'e symmetry adding 
spin 1/2 generators, the {\it supercharges}, with an algebra of anticommutators. The relevant algebra includes
commutators
\beq
\label{SUSY1}
[ M^{\mu\nu} , Q^i_\alpha ] = -{i\over4}([\sigma^\mu,\ov\sigma^\nu]Q^i)_\alpha,
\qquad\qquad
[ P_\mu , Q^i_\alpha ] = 0.
\eeq
The index $i=1,\ldots,{\cal N}$ labels the supercharges and the first relation indicates that the $Q^i_\alpha$'s 
have spin 1/2.
The superalgebra is completed by the anticommutators\footnote{We disregard the possibility of central changes
for ${\cal N}>1$.}
\beq
\label{SUSY2}
\{  {Q^i_\alpha} ,  {\ov Q^j_\dalpha} \} 
= -2i\,\delta^{ij}(\sigma^\mu)_{\alpha\dalpha}  {\partial_\mu} 
= 2\,\delta^{ij}(\sigma^\mu)_{\alpha\dalpha} {P_\mu}.
\eeq
The representations of ${\cal N}$--extended supersymmetry share several important properties:
\begin{itemize}
\item
Firstly, the 
particle states in a supermultiplet have helicities extending from a maximal $\lambda$ to $\lambda - {\cal N}/2$ (and the opposite helicities if $\lambda-{\cal N}/2 \ne - \lambda$). 
Since quantum field theory admits helicities $|\lambda|\le 1$, it admits at most ${\cal N}=4$ global supersymmetries.
For supergravity, $|\lambda|\le2$ and then ${\cal N}\le8$.
\item
Secondly, all particle states have the same mass (which can be zero) and the numbers of fermionic and bosonic 
particle states are equal, $n_B=n_F$. Similarly, for representations in terms of fields (unconstrained by a dynamical
field equation) bosonic and fermionic component fields come in equal numbers. The number of helicity zero states
is then always even. 
\item
Thirdly, for a supermultiplet of fields, the allowed interactions are strongly constrained and related. 
\item
Finally, the divergences of supersymmetric quantum field theories are much softer than in a generic case:
quadratic divergences in the scalar sector are absent and, in particular, scalar and 
Yukawa (scalar--fermion) interactions are not renormalized. 
\end{itemize}
Of course, these properties immediately indicate that supersymmetry cannot be an exact symmetry of Nature. 
Realistic theories with supersymmetry must include a mechanism of supersymmetry breaking, a condition which turns
out to be a challenge to model builders. 

It should also be mentioned that Poincar\'e supersymmetry is a limiting case of the supersymmetric extension of 
anti-de Sitter space-time symmetry $SO(2,3)$ (a contraction of the AdS superalgebra). But it is not compatible 
with the de Sitter $SO(1,4)$ algebra. This algebraic fact has important dynamical implications for supergravity theories,
which are our main subject of interest here. 

\subsection{The simplest supermultiplet and auxiliary fields} \label{secglobal}

The simplest representation of ${\cal N}=1$ supersymmetry is the chiral multiplet, which describes particle states with helicities $\pm1/2, 0,0$. It then includes a Weyl (or Majorana) spinor $\psi$ and a complex scalar $z$. But it also
introduces the concept of auxiliary fields of central importance in field representations of the supersymmetry 
algebra. The relevance of auxiliary fields, with algebraic, non-propagating field equations, is related to the 
requirement $n_B=n_F$, and the fact that counting degrees of freedom on-shell and off-shell gives different
numbers.
 
To illustrate the use of auxiliary fields, consider the sum of the free massless Klein-Gordon and 
Dirac lagrangians for a complex scalar and a Weyl (or Majorana) spinor:
\beq
\label{aux1}
{\cal L}_0 = (\partial_\mu\ov z)(\partial^\mu z) + {i\over2}\psi\sigma^\mu\partial_\mu\ov\psi 
- {i\over2}\partial_\mu\psi\sigma^\mu\ov\psi.
\eeq
It is a supersymmetric theory: under the variations
\beq
\label{aux2}
\delta z = \sqrt2 \,\epsilon\psi,
\qquad\qquad\qquad
\delta\psi_\alpha = - \sqrt2 i \, \partial_\mu z (\sigma^\mu\ov\epsilon)_\alpha,
\eeq
the lagrangian changes by a derivative and the action is then invariant. There is however trouble in the algebra.
Firstly,
\beq
\label{aux3}
[ \delta_1, \delta_2 ] z = -2i (\epsilon_2\sigma^\mu\ov\epsilon_1 - \epsilon_1\sigma^\mu\ov\epsilon_2) 
\partial_\mu z,
\eeq
which is a translation $\delta x^\mu = 
2(\epsilon_2\sigma^\mu\ov\epsilon_1 - \epsilon_1\sigma^\mu\ov\epsilon_2) = 2\,\ov\epsilon_2\gamma^\mu\epsilon_1$ 
as required by the supersymmetry algebra (\ref{SUSY2}). But 
\beq
\label{aux4}
\begin{array}{rcl}
[ \delta_1, \delta_2 ] \psi_\alpha &=& 
-2i (\epsilon_2\sigma^\mu\ov\epsilon_1 - \epsilon_1\sigma^\mu\ov\epsilon_2) \partial_\mu \psi_\alpha
\crbig 
&& + 2i(\partial_\mu\psi\sigma^\mu\ov\epsilon_2)\epsilon_{1\alpha}
- 2i(\partial_\mu\psi\sigma^\mu\ov\epsilon_1)\epsilon_{2\alpha}.
\end{array}
\eeq
The first term is as expected, but the second only vanishes if the spinor solves the Dirac equation 
$\partial_\mu\psi\sigma^\mu = 0$ implied by the lagrangian. Hence, variations (\ref{aux2}) only close the 
supersymmetry algebra for on-shell fields. This is certainly a problem if one wishes to construct 
more complicated, interacting lagrangians with nonlinear field equations. One must then simultaneously 
invent the lagrangian and the corresponding supersymmetry variations (which become nonlinear as well). 

However, modify the variation of the spinor:
\beq
\label{aux5}
\delta\psi_\alpha = - \sqrt 2 f \epsilon_\alpha - \sqrt2 i \partial_\mu z (\sigma^\mu\ov\epsilon)_\alpha
\eeq
where $f$ is a complex scalar field. The new term adds
$$
-\sqrt2 \, \delta_1f \,\epsilon_{2\alpha} + \sqrt2 \, \delta_2f \, \epsilon_{1\alpha}
$$
to $[ \delta_1, \delta_2 ] \psi_\alpha $ and choosing then
\beq
\label{aux6}
\delta f = - \sqrt 2i \, (\partial_\mu\psi\sigma^\mu\ov\epsilon)
\eeq
leads to the expected algebra
\beq
\label{aux7}
\begin{array}{rcl}
[ \delta_1, \delta_2 ] \psi_\alpha &=& 
-2i (\epsilon_2\sigma^\mu\ov\epsilon_1 - \epsilon_1\sigma^\mu\ov\epsilon_2) \,\partial_\mu \psi_\alpha ,
\crbig 
[ \delta_1, \delta_2 ] f &=& 
-2i (\epsilon_2\sigma^\mu\ov\epsilon_1 - \epsilon_1\sigma^\mu\ov\epsilon_2) \,\partial_\mu f
\end{array}
\eeq
for all three fields $z$, $\psi$ and $f$. The modification of the spinor variation also adds a new contribution
to the variation of ${\cal L}_0$:
$$
- f \delta \ov f - \ov f \delta f + {i\over\sqrt2}\partial_\mu [ f\,\epsilon\sigma^\mu\ov\psi - \ov f \,\psi\sigma^\mu\ov\epsilon ].
$$
This in turn imposes to modify the lagrangian to 
\beq
\label{aux8}
{\cal L} = (\partial_\mu\ov z)(\partial^\mu z) + {i\over2}\psi\sigma^\mu\partial_\mu\ov\psi 
- {i\over2}\partial_\mu\psi\sigma^\mu\ov\psi + \ov ff,
\eeq
with field equations 
\beq
\label{aux9}
\Box z = 0, \qquad \partial_\mu\psi\sigma^\mu = 0, \qquad f=0
\eeq
and the scalar $f$ is {\it auxiliary}: it describes $n_B=2$ off-shell fields and $n_B=0$ on-shell states. Since
$\psi$ includes $n_F=4$ off-shell and $n_F=2$ on-shell degrees of freedom while for $z$ $n_B=2$ on-shell 
and off-shell, the equality $n_B=n_F$ is verified on-shell {\it and} off-shell in the supermultiplet $(z,\psi,f)$.
On shell, $\delta f=0$ and one returns to the original expressions (\ref{aux1}) and (\ref{aux2}).

In general, the equality of the number of bosonic and fermionic physical (on-shell) degrees of freedom is 
imposed by the supersymmetry algebra while a mismatch in the numbers of bosonic and fermionic 
off-shell fields suggests that adding auxiliary fields is necessary to obtain an off-shell representation, 
if possible at all.

The canonical dimensions (in energy unit) of $z$, $\psi$ and $f$ are respectively $1$, $3/2$ and $2$ 
and the parameter $\epsilon$ has dimension $-1/2$. Hence $f$ must transform in a field with dimension
$5/2$, which is then a derivative of $\psi$. This suggests a method to construct supersymmetric lagrangians: starting with 
an off-shell supermultiplet like $(z,\psi,f)$, combine supermultiplets into a new supermultiplet 
({\it tensor calculus}) and take its component with the highest dimension as a lagrangian term: 
it necessarily transforms as a derivative. The simplest example is
\beq
\label{aux10}
Z = z^2, \qquad\qquad \Psi = 2 z\psi, \qquad\qquad F = 2fz+ \psi\psi.
\eeq
One easily verifies that $(Z,\Psi,F)$ and $(z,\psi,f)$ have identical transformations.  
Hence, since the variation of $F$ is a derivative,
\beq
\label{aux11}
\begin{array}{rcl}
{\cal L}_m &=& (\partial_\mu\ov z)(\partial^\mu z) + {i\over2}\psi\sigma^\mu\partial_\mu\ov\psi 
- {i\over2}\partial_\mu\psi\sigma^\mu\ov\psi + \ov ff
\crbig
&& - m [ fz+ {1\over2} \psi\psi ] - m[ \ov f\ov z+ {1\over2} \ov{\psi\psi} ] 
\end{array}
\eeq
is supersymmetric. Eliminating $f$ with its field equation $f=m\ov z$ leads to
\beq
\label{aux11b}
{\cal L}_m= (\partial_\mu\ov z)(\partial^\mu z) - m^2 z\ov z+ {i\over2}\psi\sigma^\mu\partial_\mu\ov\psi 
- {i\over2}\partial_\mu\psi\sigma^\mu\ov\psi - {m\over2}[ \psi\psi + \ov{\psi\psi}],
\eeq
with a common mass $m$ for $z$ and $ \psi$, and to the supersymmetry variations
\beq
\label{aux12}
\delta z = \sqrt2 \,\epsilon\psi,
\qquad\qquad\qquad
\delta\psi_\alpha = -\sqrt2 \, m\ov z \, \epsilon_\alpha - \sqrt2 i \, \partial_\mu z (\sigma^\mu\ov\epsilon)_\alpha.
\eeq
Contrary to variation (\ref{aux5}), $\delta\psi_\alpha$ now depends on the lagrangian parameter $m$.
Note that with Dirac equation $i\partial_\mu\psi\sigma^\mu = -m \ov \psi$, the on-shell variation of the auxiliary 
field is
\beq
\label{aux13}
\delta f = - \sqrt 2i \, (\partial_\mu\psi\sigma^\mu\ov\epsilon) = m\sqrt 2\,\ov{\epsilon\psi} = m\,\delta\ov z,
\eeq
as indicated by $f = m\ov z$.

Similarly, a renormalizable interaction would follow from the observation that
\beq
\label{aux14}
\begin{array}{rcl}
Z &=& {m\over2}z^2 + {\lambda\over3}z^3, \equiv W(z), \qquad\qquad\qquad 
\Psi_\alpha \,\,=\,\, (mz+\lambda z^2) \, \psi_\alpha,
\crbig  
F &=& (mz + \lambda z^2) f + {1\over2} (m+2\lambda z) \, \psi\psi
\end{array}
\eeq
is a chiral multiplet. The holomorphic function $W$ of $z$ only is the {\it superpotential}.
The supersymmetric lagrangian
\beq
\label{aux15}
\begin{array}{rcl}
{\cal L}_{m,\lambda} &=& (\partial_\mu\ov z)(\partial^\mu z) + {i\over2}\psi\sigma^\mu\partial_\mu\ov\psi 
- {i\over2}\partial_\mu\psi\sigma^\mu\ov\psi + \ov ff
\crbig
&& - (mz + \lambda z^2) f - (m\ov z + \lambda \ov z^2) f -{m\over2}[\psi\psi + \ov{\psi\psi}]
- \lambda z\psi\psi - \ov\lambda \ov z\ov{\psi\psi}
\crbig
&=& (\partial_\mu\ov z)(\partial^\mu z)  - V(z,\ov z)
\crbig
&& + {i\over2}\psi\sigma^\mu\partial_\mu\ov\psi 
- {i\over2}\partial_\mu\psi\sigma^\mu\ov\psi  - {m\over2}[ \psi\psi + \ov{\psi\psi}]
- \lambda z\psi\psi - \ov\lambda \ov z\ov{\psi\psi},
\end{array}
\eeq
using $\ov f=mz + \lambda z^2$ in the second expression, includes the scalar potential
\beq
\label{aux16}
V(z,\ov z) = |f|^2 = |mz + \lambda z^2|^2 = \left|{d\over dz}W(z)\right|^2 .
\eeq
It is a renormalizable quantum field theory, and supersymmetry holds to all orders of perturbation theory.
This result shows how supersymmetry relates all three scalar and Yukawa interactions and how the
scalar potential is related to the spinor variations
\beq
\label{aux17}
\begin{array}{l} \displaystyle
\delta \psi_{i\alpha} = - \sqrt2 \, {\cal A}_i(\varphi_i) \epsilon_\alpha + \partial_\mu(\ldots) + \ldots
\qquad
\longleftrightarrow\qquad V= \sum_i|{\cal A}_i(\varphi_i)|^2 ,
\crbig \displaystyle
{\cal A} = f = {d\ov W(\ov z)\over d\ov z} .
\end{array}
\eeq
In the first line, the index $i$ would label the various spinor and scalar fields $\varphi_i$ in the theory.
The second line refers to our example of a single chiral multiplet with superpotential $W$.
This relation between the potential and spinor variations is a universal property, even in theories and 
supermultiplets for which auxiliary fields (and then the second line) do not exist \cite{FM}. Notice that the 
on-shell spinor variation is not linear in an interacting theory.

The generalization of the "square" (\ref{aux10}) of a chiral multiplet is as follows. Consider a set of
chiral multiplets $(z^i,\psi^i, f^i)$ and an arbitrary superpotential function $W(z^i)$ of the scalar fields $z^i$. 
Then, 
\beq
\label{aux17b}
Z = W(z^i), \qquad\qquad \Psi_\alpha = {\partial W\over\partial z^i} \psi^i_\alpha, \qquad\qquad
F = {\partial W\over\partial z^i} f^i + {1\over2} {\partial^2 W\over\partial z^i\partial z^j}\psi^i\psi^j
\eeq
are the components of a chiral multiplet. Another important chiral multiplet is called {\it kinetic}. Its component 
fields are
\beq
\label{aux18}
Z = \ov f, \qquad\qquad
\Psi_\alpha = i(\sigma^\mu\partial_\mu\ov\psi)_\alpha, \qquad\qquad
F = - \Box\ov z ,
\eeq
where $(z,\psi,f)$ is a chiral multiplet.
Multiplying then the kinetic multiplet with $(z,\psi,f)$ using the tensor product rule (\ref{aux17b}) leads
to the kinetic lagrangian (\ref{aux8}):
\beq
\label{aux19}
zF + fZ + \psi\Psi = -z\Box \ov z + i \psi\sigma^\mu\partial_\mu\ov\psi + \ov ff =
{\cal L} 
+ \partial_\mu\Bigl[ {i\over2} \psi\sigma^\mu\ov\psi - z\partial^\mu\ov z \Bigr].
\eeq

The method of tensor calculus \cite{WZ} can be systematically applied to construct lagrangians invariant under 
global supersymmetry. It has found a beautiful synthesis, at least for the case of ${\cal N}=1$ supersymmetry 
(in four dimensions), in {\it superspace} and {\it superfield} techniques \cite{SS, FWZ}, 
building on the idea that supersymmetry
generators act like ``square roots of translations", as suggested by the superalgebra (\ref{SUSY2}). The 
operators $Q_\alpha$ are realized in terms of derivatives (like translations) acting in a superspace extended with 
fermionic, Grassmann (fictitious) coordinates. A tensor calculus also exists for conformal
supersymmetry (gauge theories of the superconformal algebra). It probably offers the most efficient 
procedure to construct supergravity theories, with local supersymmetry.\footnote{See section \ref{secconf}.}

A similar discussion could be made for the supermultiplet with helicities $\pm 1, \pm1/2$, which is 
realized by a gauge field $A_\mu$ and a
Majorana spinor $\lambda_\alpha$, the {\it gaugino}. Since the gaugino includes four off-shell fields while 
the gauge field has three, the off-shell supermultiplet includes one real scalar auxiliary field $D$.
Again, the Yang-Mills and Dirac lagrangians, with their non-abelian covariantizations, provide a 
supersymmetric theory: the {\it super-Yang-Mills} (SYM) lagrangian is then simply
\beq
{\cal L}_{SYM} = -{1\over4} F_{\mu\nu}^A F^{A\mu\nu} 
+ {i\over2}\lambda^A\sigma^\mu D_\mu\ov\lambda^A
- {i\over2}D_\mu\lambda^A\sigma^\mu\ov\lambda^a  + {1\over 2} D^AD^A ,
\eeq
where $D_\mu$ and $F_{\mu\nu}^a$ are the usual covariant derivative and field-strength tensor of
a non-abelian gauge theory\footnote{All fields are in the adjoint representation.},
and $D^A=0$ by its field equation. The SYM theory can be coupled to chiral multiplets in an anomaly-free
representation of the gauge group, to give the supersymmetric extension of gauge theories. 
This is the framework of the {\it minimal supersymmetric Standard Model} (MSSM), and of its variations.

It is then not surprising that the supermultiplet with helicity states $\pm2, \pm3/2$ would lead to a field theory 
combining the Einstein-Hilbert lagrangian of general relativity and the Rarita-Schwinger lagrangian for the
helicities $\pm3/2$: this leads to ${\cal N}=1$ supergravity.

\subsection{Breaking supersymmetry}

With respect to standard ``bosonic" symmetries, spontaneous breaking of supersymmetry \cite{FI, O'R}
is peculiar and difficult to achieve. Breaking a local or global symmetry 
is usually ``parameter-controlled", in the sense that the scalar potential which defines the ground state depends on 
parameters and the various possible phases correspond in general to sizeable domains in the parameter space of 
the theory. Selecting values of parameters selects the phase. The spontaneous breaking of supersymmetry 
is ``algebra-controlled": the scalar potential is a sum of 
positive terms, each term proportional to the square of an auxiliary field, either $f_i$ for chiral superfields or
$D^A$ for gauge multiplets. In the renormalizable theory,
\beq
V = \sum_i |f_i|^2 + {1\over2} \sum_A D^AD^A.
\eeq
By their algebraic field equations, the auxiliary fields are functions of the chiral 
scalars $z^i$ and if equations 
\beq
f_i (z_i) = D^A(z_i,\ov z_i) =0
\eeq
have a solution, this solution is the true ground state of the theory and supersymmetry is not broken.\footnote{ 
The potential vanishes then at a supersymmetric minimum. But since general relativity is absent, the value
of the potential at the ground state, sometimes called {\it vacuum energy}, does not have any physical significance.}
We then have
an algebraic condition for supersymmetry breaking, that these equations cannot be solved. 
 
Spontaneous supersymmetry breaking has two undesired consequences. Firstly, it generates a 
massless spin 1/2 Goldstone particle, the Goldstino. This can be seen for instance in the variation of 
$\psi_\alpha$, eq.~(\ref{aux5}). If $f$ acquires a vacuum expectation value $\langle f\rangle$
\beq
\delta \psi_\alpha = \sqrt2 \langle f \rangle \epsilon_\alpha + \ldots
\eeq
and the inhomogeneous term is typical of Goldstone particles. Secondly, if spontaneous supersymmetry 
breaking is able to lift fermion--boson mass degeneracies, it in general moves the mass of some spin zero states below 
their fermionic partner, in contradiction with observations. These obstructions can be avoided if the spontaneously broken supersymmetry is local, {\it i.e.} in a theory with gauged supersymmetry. This is a first motivation, from
a low-energy perspective, for supergravity, with the idea that the residual, effective effects of the breaking will 
produce the mass terms necessary in a realistic particle spectrum. Models realising this idea are actually easy to construct, and they are at the origin of the various supersymmetric extensions of the Standard Model under 
test at LHC experiments.

\section{Supergravity} \label{local}
\setcounter{equation}{0}

Supergravity is the theory of gauged supersymmetry. The spin 1/2 parameter $\epsilon_\alpha$ is then local,
$\epsilon_\alpha(x)$, and since translation generators $P_\mu$ appear in the supersymmetry algebra 
(\ref{SUSY2}), translations are local as well. This calls for coordinate diffeomorphisms (general coordinate transformations, GCT) and then for general relativity and gravitation. The theory of gauged supersymmetry is 
then a field theory of gravitation.

A local symmetry requires a gauge field (a connection) to construct tensors and invariants involving derivatives,  
needed in lagrangians and dynamical field equations. Since supercharges $Q_\alpha$ 
and parameters $\epsilon_\alpha$ are 
Lorentz spinors, the gauge field of supersymmetry is a vector-spinor field, $\psi_{\alpha\mu}$, the {\it gravitino}, 
and the physical (massless) states will have helicities $\pm3/2$.
To construct a supersymmetric lagrangian, we first need a kinetic lagrangian for the gravitino: 
the Rarita-Schwinger lagrangian. We may then add this term to the Einstein-Hilbert lagrangian for the metric 
tensor and maybe find supersymmetry variations leaving the action invariant and closing the supersymmetry 
algebra for solutions of the field equations, on-shell. 

Or we may try to directly obtain an off-shell representation of supersymmetry including the metric tensor, the 
gravitino and, if needed, auxiliary fields. Let us count off-shell degrees of freedom:
\begin{itemize}
\item
The metric tensor $g_{\mu\nu}$ has ten components, four can be removed by gauge transformations 
(local translations) to remain with six fields, $6_B$.
\item
The gravitino is a Majorana vector--spinor with four local gauge supersymmetries. It includes then
$4\times 4-4=12_F$ component fields.\footnote{Alternatively, the supercharge $Q_\alpha$ includes four 
operators and one (fermionic) gauge field ($3_F$) is needed for each of them.}
\item
With $6_B+12_F$ propagating fields, $(6+n)_B + n_F$ non-propagating auxiliary fields are then needed 
to construct an off-shell representation.
\end{itemize}
It turns out that several possibilities exist.\footnote{For a review and a comparison of different choices 
in the superconformal approach, see \cite{FGKVP}.}
Minimal supergravities have $6_B$ auxiliary fields ($n=0$).
{\it Old minimal} supergravity \cite{StW, FPvN} has a complex scalar ($2_B$) and a vector field not associated with 
a gauge symmetry ($4_B$), {\it new minimal} supergravity \cite{SoW} has an antisymmetric tensor $B_{\mu\nu}$
and a vector field $A_\mu$ with gauge symmetries
\beq
\delta B_{\mu\nu} = \partial_\mu\Lambda_\nu - \partial_\nu\Lambda_\mu \qquad (6-3)_B, \qquad\qquad
\delta A_\mu = \partial_\mu\Lambda \qquad (4-1)_B.
\eeq
And there are {\it non-minimal} versions with $n\ne 0$, including some versions which have supplementary
propagating fields.\footnote{``$16_B+16_F$" supergravity \cite{16A, 16B, 16C} for instance is related to string theory compactifications \cite{16D},
or to supercurrent structures \cite{KS, ADH}.} In pure supergravity, with $\pm3/2$ and $\pm2$ physical states,
the formulation does not matter much since auxiliary fields anyway vanish. They play however a role when 
coupling supergravity to other supermultiplets: they define classes of admissible interactions which depend 
directly on the choice of supergravity auxiliary fields \cite{FGKVP}.

\subsection{Spinors and the vierbein}

From here on, $x^\mu$ denotes coordinates of a space-time with metric tensor $g_{\mu\nu}(x)$ and line element
$ds^2 = g_{\mu\nu}(x)dx^\mu dx^\nu$.

Spinor fields live in the flat minkowskian space tangent at each point $x$. This tangent space would 
have local coordinates $\zeta^a(x)$ with line element\footnote{$\eta_{ab}={\rm diag}(1,-1,-1,-1)$ is the flat
Minkowski metric.}
\beq
\label{ds}
ds^2 = \eta_{ab} \, d\zeta^a d\zeta^b = \eta_{ab}(\partial_\mu\zeta^a) (\partial_\nu\zeta^b) dx^\mu dx^\nu.
\eeq
Hence, the transition from the curved space-time to the flat tangent space is given by the sixteen
fields $e_\mu^a(x) = \partial_\mu\zeta^a(x)$, in other words, we can define a {\it vierbein} $e_\mu^a$
and its inverse $e^\mu_a$ (since the metric has an inverse) 
such that
\beq
\label{bein1}
g_{\mu\nu} = \eta_{ab} \, e_\mu^a e_\nu^b, \qquad\qquad  e^\mu_a e^a_\nu = \delta^\mu_\nu, \qquad
e^a_\mu e_b^\mu = \delta ^a_b.
\eeq
At each point $x$, a Lorentz algebra acts in the tangent space. This local Lorentz symmetry 
allows to eliminate six components of the vierbein and since $ds^2$ in (\ref{ds}) is Lorentz invariant, the ten 
remaining components are the ten components of $g_{\mu\nu}$. It also acts on spinors:
\beq
\delta \psi(x) = {1\over2} \omega_{ab}\, \sigma^{ab} \psi (x) ,
\qquad\qquad \sigma^{ab}={1\over2}\gamma^{ab}, \qquad
\gamma^{ab} = {1\over2}[ \gamma^a,\gamma^b ].
\eeq
Covariant derivatives of spinors are provided by the local Lorentz gauge field, the {\it spin connection} 
${\omega_\mu}^{ab} = -{\omega_\mu}^{ba}$:
\beq
D_\mu \psi = \partial_\mu\psi + {1\over2}\omega_{\mu\,ab}\, \sigma^{ab} \psi
\eeq
and the Dirac lagrangian takes then the form
\beq
e^{-1}{\cal L}_\psi = i\ov\psi\gamma^\mu D_\mu \psi  
\eeq
where $e=\sqrt{|\det g_{\mu\nu|}} = \det e_\mu^a$ and $\gamma^\mu = e^\mu_a \gamma^a$. It is invariant
under GCT and local Lorentz.

\subsection{The gravitino and the Rarita-Schwinger action}
\label{secRarita}

The Rarita-Schwinger action describes the propagation of a vector-spinor
field $\psi_{\alpha\mu}$ in the background defined by the vierbein $e_\mu^a$ 
and the spin connection ${\omega_\mu}^{ab}$. Its form is dictated by 
invariance requirements and reduction to the relevant helicity components only.
Under the Lorentz algebra, the field $\psi_{\alpha a} = e^\mu_a\psi_{\alpha \mu}$ ($\alpha$ is a spinor index)
transforms in the reducible representation
$$
\begin{array}{ccccccc}
{\rm spinor}&\otimes&{\rm vector} &=& {\rm gravitino}&\oplus&{\rm spinor} \,,
\cr
[(2,1)\oplus(1,2)]&\otimes& (2,2) &=& [(3,2)\oplus(2,3)]
&\oplus&[(1,2)\oplus(2,1)]\,.
\end{array}
$$
The second equation indicates the representations of the Lorentz algebra
$SO(1,3) \sim Sl(2,\mathbb{C})$, the numbers are the dimensions of $Sl(2,\mathbb{R})$ 
representations. The spinor part of $\psi_{\alpha a}$ is $\gamma^a\psi_{\alpha a}$ and the gravitino part is 
then isolated by the condition
\beq
\label{proj}
(\gamma^a\psi_a)_\alpha = (\gamma^\mu\psi_\mu)_\alpha = 0
\qquad\Longrightarrow\qquad
\widetilde\psi_{\alpha a} = \psi_{\alpha a} - {1\over4}(\gamma_a\gamma^b\psi_b)_\alpha.
\eeq 
An action for the gravitino should in principle include this projection condition in its field equations.

Consider then the following free lagrangian density, in Minkowski space
(coordinates $\zeta^a$ and metric $\eta_{ab}$):
\beq
\label{freeRS1}
{\cal L}_0 = {1\over2\kappa^2} \, \ov\psi_a\gamma^{abc}\partial_b\psi_c
\eeq
where $\kappa$ is a constant with dimension (mass)$^{-1}$ and $\gamma^{abc} 
= \gamma^{[a}\gamma^b\gamma^{c]} =
{1\over6}\gamma^a\gamma^b\gamma^c \pm 5$ terms. 
The gravitino $\psi_a$ is Majorana and ${\cal L}_0$ is hermitian.
It implies the field equation
\beq
\label{freeRS2}
\gamma^{abc}\partial_b\psi_c=0.
\eeq
Invariance under the gauge transformation $\delta\psi_a = \partial_a\lambda$, with an arbitrary Majorana spinor
$\lambda$, can be used to impose the projection condition (\ref{proj}) by solving
$\gamma^a\partial_a\lambda = -\gamma^a\psi_a$. This leaves a residual gauge symmetry 
$\delta\psi_a = \partial_a\tilde\lambda$ with $\tilde\lambda$ solution of the massless Dirac equation 
$\gamma^a\partial_a\tilde\lambda=0$. In the gauge $\gamma^a\psi_a=0$, the field equation reduces to
\beq
\label{freeRS3}
\gamma^a \partial_b\psi^b = \gamma^b\partial_b\psi^a
\eeq
and multiplication by $\gamma_a$ leads to
\beq
\gamma^a\psi_a=0 \quad \makebox{(gauge choice)}, \qquad\qquad
\partial_b\psi^b = 0, \qquad\qquad \gamma^b\partial_b\psi_a=0 \qquad\makebox{(Dirac)}.
\eeq
The Dirac equation indicates that the field is massless and the count of physical degrees of freedom is as follows.
Starting with $16_F$ fields, the gauge choice and $\partial^a\psi_a=0$ remove two spinors ($8_F$), the
massless Dirac equation removes four of the $8_F$ remaining fields and finally the residual gauge symmetry eliminates one of the massless Dirac spinor $(2_F$) to leave only two degrees of freedom, which turn out to have helicities $\pm3/2$.\,\footnote{Plane waves $\epsilon_{\alpha a}(k) e^{-ikx}$ can be used to see this.}

Coupling theory (\ref{freeRS1}) to the background described by the vierbein $e_\nu^a$ leads to the Rarita-Schwinger
lagrangian
\beq
\label{RS1}
e^{-1} {\cal L}_{RS} = {1\over2\kappa^2} \, \ov\psi_\mu\gamma^{\mu\nu\rho}\widetilde D_\nu\psi_\rho,
\eeq
where $\gamma^{\mu\nu\rho} = e^\mu_a e^\nu_b e^\rho_c \gamma^{abc}$. In principle, since the gravitino field
$\psi_{\alpha\mu}$ is a space-time vector (index $\mu$), its covariant derivative should be
\beq
\label{psimuder}
D_\mu \psi_{\alpha\rho} = \partial_\mu\psi_{\alpha\rho} - \Gamma^\sigma_{\mu\rho}(g)\psi_{\alpha\sigma}
+ {1\over2} {\omega_\mu}^{ab} ( \sigma_{ab}\psi_\rho)_\alpha
\eeq
with affine connection
\beq
\label{Gammais}
\Gamma^\mu_{\nu\rho}(g) = {1\over2}g^{\mu\sigma} [
\partial_\nu g_{\rho\sigma} + \partial_\rho g_{\nu\sigma} - \partial_\sigma g_{\nu\rho} ].
\eeq
But the antisymmetry of $\gamma^{\mu\nu\rho}$ removes the symmetric affine connection and
\beq
\label{psimuder2}
\widetilde D_\mu \psi_\nu = \partial_\mu\psi_\nu 
+ {1\over2} {\omega_\mu}^{ab}\, \sigma_{ab}\psi_\nu
\eeq
appears in the Rarita-Schwinger lagrangian (\ref{RS1}).

\subsection{Simple ${\cal N}=1$ supergravity}

The need for spinor fields imposes a formulation of general relativity in terms of the vierbein. Since the Rarita-Schwinger also uses the spin connection, it is natural to use the {\it first order} (or Palatini) formalism in which
$e_\mu^a$ and ${\omega_\mu}^{ab}$ are independent fields, their relation being a field equation. One then
introduces the curvature tensor of the spin connection,
\beq
\label{Lcurvature}
{R_{\mu\nu}}^{ab} = \partial_\mu {\omega_\nu}^{ab} 
-\partial_\nu {\omega_\mu}^{ab} + {\omega_\mu}^{ac}{\omega_{\nu\,c}}^b
- {\omega_\nu}^{ac}{\omega_{\mu\,c}}^b = -{R_{\nu\mu}}^{ab}  =
-{R_{\mu\nu}}^{ba}\,,
\eeq
the curvature scalar
\beq
\label{Ris1}
R = {R_{\mu\nu}}^{ab}e^\mu_ae^\nu_b,
\eeq
and the gravity lagrangian
\beq
\label{EHaction}
{\cal L}_{grav.} = {1\over2\kappa^2}\, eR \,.
\eeq
All quantities are diffeomorphism and Lorentz tensors or scalars and the gravitational coupling constant is 
$\kappa= \sqrt{8\pi}M_P^{-1}$ in terms of the Planck scale $M_P\simeq1.2\times 10^{19}$ GeV. 
Under a variation of the vierbein,
\beq
\label{Sgravvar1}
\delta e^\mu_a {\delta\over\delta e^\mu_a} {\cal L}_{grav.} = 
{1\over\kappa^2} e \left[ {R_{\mu\nu}}^{ab}e_b^\nu
-{1\over2}e_\mu^a R\right]\delta e^\mu_a
\eeq
leads to Einstein equation, after the elimination of the spin connection.

Since the action (\ref{EHaction}) is quadratic in the spin connection
and linear in its first derivative, the Euler-Lagrange equation
for ${\omega_\mu}^{ab}$ is algebraic only. To calculate this field equation, rewrite
\beq
\begin{array}{rcl}
eR &=& e(e^\mu_a e^\nu_b-e^\mu_b e^\nu_a)
\left( \partial_\mu{\omega_\nu}^{ab}
+{\omega_{\mu}}^{ac}{{\omega_{\nu}}_c}^b\right) \crbig
&=&
- {\omega_\nu}^{ab} \partial_\mu[e(e^\mu_a e^\nu_b-e^\mu_b e^\nu_a)] 
+e(e^\mu_a e^\nu_b-e^\mu_b e^\nu_a)
{\omega_{\mu}}^{ac}{{\omega_{\nu}}_c}^b +{\rm derivative}
\end{array}
\eeq
and the field equation leads to 
\beq
\label{omegais1}
\begin{array}{rcl}
\omega_{\mu\,cd} &=&
-{1\over2}(\partial_\mu e_{\nu c}- \partial_\nu e_{\mu c}) e^\nu_d
+{1\over2}(\partial_\mu e_{\nu d}- \partial_\nu e_{\mu d}) e^\nu_c
-{1\over2}e^\rho_c e^\nu_d(\partial_\rho e_{\nu a} -
\partial_\nu e_{\rho a})e^a_\mu \crbig
&\equiv& \omega_{\mu\,cd}(e).
\end{array}
\eeq

In terms of $g_{\mu\nu}$ and of the symmetric $\Gamma^\lambda_{\mu\nu}(g)$ (\ref{Gammais}), 
the Ricci tensor $R_{\mu\nu}$ corresponding to definitions (\ref{Lcurvature}) and (\ref{Ris1}) is
\beq
\label{RChristoffel}
R_{\mu\nu} = \partial_\rho\Gamma^\rho_{\mu\nu}(g)
- \partial_\nu\Gamma^\rho_{\rho\mu}(g)
-\Gamma^\rho_{\sigma\mu}(g)\Gamma^\sigma_{\rho\nu}(g)
+\Gamma^\rho_{\mu\nu}(g)\Gamma^\sigma_{\rho\sigma}(g) \,\, = \,\,
R_{\nu\mu} 
\eeq
and $R = g^{\mu\nu}R_{\mu\nu}$.

Next, we combine the gravity and Rarita-Schwinger lagrangians:
\beq
\label{ERS1}
{\cal L}_{ERS}( e^a_\mu,\psi_{\alpha\mu},\omega_{\mu\,ab}) 
= {1\over2\kappa^2}\,e\left(R
+\ov\psi_\mu\gamma^{\mu\nu\rho} {\tilde D_\nu\psi_\rho}\right).
\eeq
After elimination of the spin connection, using its algebraic field equation, we will obtain an interacting 
theory for the propagating vierbein and gravitino. Since the gravitino lagrangian also includes a term linear 
in ${\omega_\mu}^{ab}$ in the Lorentz covariant derivative $\widetilde D_\mu$,
its field equation and its solution are modified. As a consequence, the spin connection acquires {\it contorsion},
\beq
\label{cont1}
{\omega_\mu}^{ab} = {\omega_\mu}^{ab} (e) + {\kappa_\mu}^{ab},
\eeq
and the contorsion tensor is quadratic in the gravitino field:
\beq
\label{cont2}
\kappa_{\mu\,ab} = -{1\over4}\left[ \ov\psi_\mu\gamma_a\psi_b - \ov\psi_\mu\gamma_b\psi_a 
+ \ov\psi_a\gamma_\mu\psi_b \right] = -\kappa_{\mu\,ba}.
\eeq
In the Rarita-Schwinger lagrangian,
\beq
\tilde D_\mu \psi_\nu - \tilde D_\nu \psi_\mu = 
\widehat D_\mu\psi_\nu - \widehat D_\nu\psi_\mu + 2S^\lambda_{\mu\nu}\psi_\lambda,
\qquad\qquad
\widehat D_\mu\psi_\nu = \partial_\mu\psi_\nu  + {1\over2}{\omega_\mu}^{ab} (e) \sigma_{ab}\psi_\nu,
\eeq
with torsion tensor
\beq
S^\lambda_{\mu\nu} = -{1\over4}
\ov\psi_\mu\gamma^\lambda\psi_\nu 
\eeq
defined as the antisymmetric part of the affine connection
$\Gamma^\mu_{\nu\rho} = \Gamma^\mu_{\nu\rho} (g) + S^\mu_{\nu\rho}$, 
$S^\mu_{\nu\rho} = -S^\mu_{\rho\nu}$.

Useful formulas are obtained by inserting the decomposition (\ref{cont1}) into the gravity lagrangian, after 
some partial integrations:
\beq
\label{eRaction1}
\begin{array}{rcl}
eR &=& eR \left(\omega(e)\right) - e[{\kappa^a}_{ac}{\kappa_b}^{bc} 
- \kappa_{abc}\kappa^{cba}] + \,\,{\rm derivative} \,, \crbig
eR\left(\omega(e)\right) &=& e[{\omega^a}_{ac}(e){\omega_b}^{bc}(e) 
- \omega_{abc}(e)\omega^{cba}(e)] + \,\,{\rm derivative} \,,
\end{array}
\eeq
where $\omega_{abc}(e) = e^\mu_a\,\omega_{\mu\,bc}(e)$ and 
$\kappa_{abc} = e^\mu_a\,\kappa_{\mu\,bc}$ and the derivatives can be dropped in the lagrangian. 

Inserting the spin connection (\ref{cont1}) with contorsion (\ref{cont2}) in ${\cal L}_{ERS}$
leads to the lagrangian density of ${\cal N}=1$ pure supergravity, as a function of $e_\mu^a$ 
and $\psi_\mu$ only ({\it second order formalism}):
\beq
\label{2ndaction}
\begin{array}{rcl}
{\cal L} &=& \displaystyle {1\over2\kappa^2}eR\left(\omega(e)\right)
+ {1\over2\kappa^2} e \ov\psi_\mu\gamma^{\mu\nu\rho}\widehat D_\nu\psi_\rho
\crbig
&& \displaystyle +{e\over32\kappa^2}\left[ 4(\ov\psi^\mu\gamma_\mu\psi_\rho)
(\ov\psi^\nu\gamma_\nu\psi^\rho)
-(\ov\psi_\mu\gamma_\nu\psi_\rho)(\ov\psi^\mu\gamma^\nu\psi^\rho)
-2(\ov\psi_\mu\gamma_\nu\psi_\rho)(\ov\psi^\mu\gamma^\rho\psi^\nu)\right],
\end{array}
\eeq
with now $\widehat D_\nu\psi_\rho = \partial_\nu\psi_\rho
+{1\over2}\omega_{\nu\,ab}(e)\sigma^{ab}\psi_\rho$. 

With some efforts,\footnote{Using properties of Majorana
spinors, Fierz rearrangements and partial integrations.} one can show that ${\cal L}$ transforms with a 
derivative under the local supersymmetry variations
\beq
\label{4dsusy1}
\begin{array}{rclrcl}
\delta e_\mu^a &=& -{1\over2}\ov\epsilon\gamma^a\psi_\mu \,, \qquad&
\delta e^\mu_a &=& {1\over2}\ov\epsilon\gamma^\mu\psi_a \,, \crbig
\delta\psi_\mu &=& D_\mu\epsilon \,\,=\,\,
\partial_\mu\epsilon +{1\over2}\omega_{\mu\,ab} \, \sigma^{ab}\epsilon \,,
\qquad &
\delta\ov\psi_\mu &=& D_\mu\ov\epsilon \,\,=\,\,
\partial_\mu\ov\epsilon -{1\over2}\omega_{\mu\,ab}\,\ov\epsilon\sigma^{ab} \,,
\end{array}
\eeq
in the first-order formalism. Using eqs.~(\ref{cont1}) and (\ref{cont2}), the gravitino supersymmetry variation 
acquires a fermionic nonlinear contribution through the contorsion tensor:
\beq
\label{2ndtransf}
\delta\psi_\mu = \widehat D_\mu\epsilon -{1\over8}
\left(2\ov\psi_\mu\gamma_a\psi_b + \ov\psi_a\gamma_\mu\psi_b\right)\sigma^{ab}
\epsilon \,, 
\qquad\qquad
\widehat D_\mu\epsilon = \partial_\mu\epsilon
+{1\over2}\omega_{\mu\,ab}(e) \, \sigma^{ab}\epsilon.
\eeq
The gravitino transforms as the gauge field of supersymmetry: the first term is the derivative of the transformation
parameter $\epsilon$.

One easily obtains the algebra
\beq
\label{4dsusyalg}
[\delta_1,\delta_2]\, e_\mu^a = 
\delta_1\bigl[-{1\over2}\ov\epsilon_2\gamma^a\psi_\mu\bigr]
- \delta_2\bigl[-{1\over2}\ov\epsilon_1\gamma^a\psi_\mu\bigr]
= -{1\over2}\,D_\mu(\ov\epsilon_2\gamma^a\epsilon_1) \,,
\eeq
the covariant derivative acting on the Lorentz vector 
$\ov\epsilon_2\gamma^a\epsilon_1$:
$$
D_\mu(\ov\epsilon_2\gamma^a\epsilon_1) = 
\partial_\mu(\ov\epsilon_2\gamma^a\epsilon_1)
+{\omega_\mu}^{ab}(\ov\epsilon_2\gamma_b\epsilon_1) \,.
$$
The quantity $\xi^\mu = e^\mu_a (\ov\epsilon_2\gamma^a\epsilon_1)$ 
is then the parameter of the infinitesimal coordinate trans\-for\-mation
predicted by the supersymmetry algebra. 
But as earlier mentioned, we do not expect with $6_B+12_F$ off-shell fields 
that the supersymmetry algebra closes without the field equations. Auxiliary fields would be needed and,
in the case of ${\cal N}=1$ supergravity, exist and are not unique.

\subsection{Anti-de Sitter supergravity: the cosmological constant}\label{seccosmo}

The natural background geometry of the supergravity lagrangian (\ref{2ndaction}) is flat Min\-kow\-ski
space. Since the matter energy-momentum tensor is entirely generated by the gravitino, it vanishes in
a Lorentz-invariant background.\footnote{There could be gravitino condensates \cite{EM}, recently 
reviewed in \cite{M}.} 
But we know that Poincar\'e supersymmetry is a limit case of the 
more general supersymmetry in anti-de Sitter (AdS) space-time. A supergravity theory with natural AdS 
background geometry should then exist. 

We use the following standard definition of the cosmological constant $\Lambda$: 
it should contribute to Einstein equations as
\beq
\label{cosmo2}
\begin{array}{rcl}
R_{\mu\nu}^{ab} e^\nu_b - {1\over2}e_\mu^a R &=& -\Lambda e^a_\mu +
\hbox{contributions from other fields}, 
\crbig
R &=& 4\Lambda + \hbox{contributions from other fields}. 
\end{array}
\eeq
A positive (negative) $\Lambda$ 
leads to de Sitter (anti-de Sitter) space-time. The field equation 
(\ref{cosmo2}) follows from the lagrangian density
\beq
\label{cosmo4}
{e\over\kappa^2}\Bigl[{1\over2}R - \Lambda \Bigr] .
\eeq

The situation is somewhat similar to the introduction of mass in the chiral supermultiplet theory discussed in 
section \ref{secglobal}. Keeping the supersymmetry variation $\delta e_\mu^a$ unchanged, the gravitino 
variation is modified to
\beq
\label{cosmo6}
\delta \psi_\mu = D_\mu\epsilon - {1\over2}M\,\gamma_\mu\epsilon \,,
\eeq
with a real number $M$.\footnote{Reality follows from the Majorana property of $\psi_\mu$.}
The modified variation of the gravitino kinetic term requires the presence in ${\cal L}$ of a quadratic, mass-like 
term for the gravitino, and the variation of this term implies the existence of a negative cosmological constant
proportional to $M^2$. The resulting lagrangian is then:\footnote{In terms of ${\omega_\mu}^{ab} 
= {\omega_\mu}^{ab}(e) + {\kappa_\mu}^{ab}$.}
\beq
\label{cosmo8}
{\cal L}_{AdS} = {e\over\kappa^2}\left[ {1\over2}R +{1\over2}\ov\psi_\mu
\gamma^{\mu\nu\rho}\tilde D_\nu\psi_\rho + 
{M\over2}\ov\psi_\mu\gamma^{\mu\nu}\psi_\nu + 3M^2\right] .
\eeq
The gravitino mass-like term 
in the lagrangian density (\ref{cosmo8}) does not mean that the gravitino 
is massive: the theory is supersymmetric, the graviton is massless, the gravitino must then be massless.
Actually, the cosmological constant $\Lambda = -3M^2$ in Einstein equation (\ref{cosmo2}) 
propagates graviton waves along light-like curves in the anti-de Sitter 
geometry. Similarly, the gravitino mass-like term is precisely
the contribution required to propagate gravitino waves on light-like 
curves in this geometry. Again, we find that a positive cosmological constant is not compatible with 
supersymmetry, as the basic superalgebra already indicates.

\subsection{The superconformal derivation, old minimal supergravity} \label{secconf}

In section \ref{secglobal}, we have constructed the (globally) supersymmetric theory of a chiral multiplet $(z,\psi,f)$, 
to illustrate the role of auxiliary fields. In this paragraph, we outline a similar approach, in the context of 
superconformal symmetry, to construct ${\cal N}=1$ supergravity with the old minimal set of auxiliary fields.
The reason to consider this construction here is that this procedure generalizes very well to theories describing 
generic ${\cal N}=1$ supersymmetric gauge theories 
coupled to supergravity \cite{CFGVP, FVP}, however at the price of a considerable increase in technical
complexity. 

The Poincar\'e and Anti-de Sitter ${\cal N}=1$ superalgebras are subalgebras of the ${\cal N}=1$ 
superconformal algebra $SU(2,2|1)$. Its bosonic sector $SU(2,2) \times U(1)_R$ includes the conformal algebra
$SU(2,2)\sim SO(2,4)$ and $U(1)_R$ symmetry, and $SO(2,4)\supset SO(1,3)\times SO(1,1)$, where
$SO(1,3)$ is Lorentz algebra and $SO(1,1)$ generates dilatation or scale or Weyl transformations. A supermultiplet
of Poincar\'e supersymmetry is also a representation of the superconformal algebra once a $SO(1,1)$ Weyl 
weight $w$ and a $U(1)_R$ chiral charge $q$ have been assigned to all fields and with the appropriate
symmetry and supersymmetry variations. There are restrictions on these 
quantum numbers. For instance $w=q$ for (the lowest component of) a chiral multiplet.\footnote{Two 
conventions, different by the 
normalization of $U(1)_R$, exist in the literature: either $w=q$ as used here, or $q={2\over3}w$.}

To construct ${\cal N}=1$ Poincar\'e supergravity, one uses a chiral supermultiplet $S_0$ with Weyl and 
$U(1)_R$ weights 
$w=q=1$ and component fields $z_0$ ($w=q=1$), $\psi_\alpha$ ($w=3/2$, $q=-1/2$) and $f_0$ ($w=2$, 
$q=-2$).\footnote{The weights of a supermultiplet are the weights of its ``lowest" component, in our cse $z_0$.
Apart from minor differences (metric sign, two-component spinors), we use the notation 
of ref.~\cite{KU}.} Its conjugate antichiral $\ov S_0$, with weights $w=-q=1$ has components $\ov z_0$ 
($w=-q=1$), $\ov\psi_\dalpha$ ($w=3/2$, $q=1/2$) and $\ov f_0$ ($w=2$, $q=2$). 
The chiral kinetic multiplet of $S_0$, denoted by $T(S_0)$, analogous to expressions (\ref{aux18}), has components
\beq
\begin{array}{rcll}
\label{conf1}
Z_0 &=& \ov f_0 \qquad &(w=q=2),
\crbig
\Psi_{0\,\alpha} &=& i(\sigma^\mu D_\mu^{\cal C} \, \ov\psi_0)_\alpha \qquad\qquad &(w=5/2, \,\, q=1/2),
\crbig
F_0 &=& -\Box^{\cal C} \ov z_0 \qquad &(w=3, \,\, q=-1).
\end{array}
\eeq
The derivatives $D_\mu^{\cal C}$ and $\Box^{\cal C}$ are covariant under the full local superconformal algebra.
Gauge fields are:
\beq
\label{conf2}
\begin{array}{llll}
e_\mu^a \qquad &\makebox{(vierbein, translations)}, \qquad
&{\omega_\mu}^{ab} &\makebox{(spin connection, Lorentz)}, 
\crbig
\psi_{\alpha\mu} &\makebox{(gravitino, supersymmetry)}, &&
\crbig
f_\mu^a &\makebox{(conformal boosts)}, \qquad
&\phi_{\alpha\mu} \qquad &\makebox{(special supersymmetry)}, 
\crbig
b_\mu \qquad &\makebox{(dilatation)}, \qquad
& A_\mu \qquad &\makebox{($U(1)_R$)}.
\end{array}
\eeq
Constraints lead to algebraic expressions for ${\omega_\mu}^{ab}$ (as earlier), $\phi_{\alpha\mu}$ and $f_\mu^a$,
leaving bosonic gauge fields $e_\mu^a$, $b_\mu$, $A_\mu$ ($6+3+3=12_B$) and the gravitino
$\psi_{\alpha\mu}$ ($12_F$).
The idea is then to write  a superconformal lagrangian for the chiral multiplet $S_0$ and then to reduce the symmetry 
to local Poincar\'e symmetries, by applying appropriate gauge fixing conditions for conformal boosts, dilatation,
$U(1)_R$ and special supersymmetry. These conditions assign values to
\beq
\label{}
\begin{array}{ll}
z_0: \qquad &\makebox{dilatation and $U(1)_R$, (modulus and phase of $z_0$),}
\crbig
\psi_0: &\makebox{special supersymmetry,}
\crbig
b_\mu: &\makebox{conformal boosts.}
\end{array}
\eeq
We are then left with the propagating fields of Poincar\'e supergravity, $e_\mu^a$ and $\psi_{\alpha\mu}$ and
the auxiliary field $A_\mu$ (the gauge field of gauge-fixed $U(1)_R$) and $f_0$ (in $S_0$).

In global supersymmetry, we obtain invariant lagrangians by combining supermultiplets into other supermultiplets 
(tensor calculus), or by multiplying superfields, and by selecting the highest-dimensional component which 
transforms with a derivative. In the superconformal case, there are two (related) possibilities to produce 
invariant action terms. Firstly, we can
combine supermultiplets to obtain a real supermultiplet with weights $w=2$, $q=0$ and take the 
{\it $D$--density formula}. For instance, symbolically,
$$
\Bigl[ S_0\ov S_0 \Bigr]_D.
$$
Secondly, we can combine chiral multiplets into another chiral multiplet with weights $w=q=3$ and 
use the {\it $F$--density formula}. In our case,
$$
\Bigl[ S_0 \, T(S_0) \Bigr]_F \qquad\qquad {\rm or}\qquad\qquad \Bigl[ S_0^3 \Bigr]_F.
$$
Up to conventions (and partial integration), $[S_0\ov S_0]_D$ and $[S_0T(S_0)]_F$ are equivalent. 

To obtain the ${\cal N}=1$ Poincar\'e supergravity, start then with the superconformal lagrangian
\beq
{\cal L} = -{3\over2} \Bigl[ S_0 \ov S_0 \Bigr]_D + \lambda \Bigl[ S_0^3 \Bigr]_F
\eeq
calculated using superconformal tensor calculus and the density formulas \cite{KU}.
Applying the gauge fixing conditions
\beq
b_\mu = 0, \qquad\qquad \psi_0 = 0,
\eeq
but retaining $z_0$ for a moment, the superconformal lagrangian reads
\beq
\begin{array}{rcl}
e^{-1} {\cal L} &=& {1\over2} z_0\ov z_0 
\Bigl[ R\left(\omega(e)\right)
+ \ov\psi_\mu\gamma^{\mu\nu\rho}\widehat D_\nu\psi_\rho \Bigr]
\crbig
&&
-{3\over2}\Bigl [ 2(D_\mu\ov z_0)(D^\mu z_0) + 2 f_0\ov f_0  \Bigr]
+ \lambda \Bigl[ 3z_0^2 f_0 + 3\ov z_0^2\ov f_0 \Bigr] + \ldots
\end{array}
\eeq
The covariant derivative is $D_\mu z_0 = \partial_\mu z_0 - {i\over2} q A_\mu z_0$ ($q=1$) and some gravitino 
interactions have been omitted. With the dilatation and $U(1)_R$ gauge choice $z_0=\kappa^{-1}$,
\beq
\begin{array}{rcl}
e^{-1} {\cal L} &=& \displaystyle {1\over2\kappa^2} 
\Bigl[ R\left(\omega(e)\right)
+ \ov\psi_\mu\gamma^{\mu\nu\rho}\widehat D_\nu\psi_\rho \Bigr]
- {3\over4\kappa^2} A_\mu A^\mu
- 3 f_0\ov f_0 
\crbig
&& \displaystyle
+ \lambda \Bigl[ 3z_0^2 f_0 + 3\ov z_0^2\ov f_0 \Bigr] + \ldots
\end{array}
\eeq
The first line displays the auxiliary fields $A_\mu$ and $f_0$ of old minimal supergravity.
Eliminating them
leads finally to
\beq
e^{-1} {\cal L} = {1\over\kappa^2} 
\left[ {1\over2} R\left(\omega(e)\right)
+ {1\over2} \ov\psi_\mu\gamma^{\mu\nu\rho}\widehat D_\nu\psi_\rho 
+ 3 {\lambda^2\over\kappa^2} \right] + \ldots ,
\eeq
with an Anti-de Sitter cosmological constant $\Lambda = -3\lambda^2\kappa^{-2}$ induced by the $F$--density.
The related gravitino mass-like term required by supersymmetry\footnote{As in eq.~(\ref{cosmo8}).} is 
actually generated by the omitted gravitino term
$$
{\lambda\over4} \, z_0^3 \, \ov\psi_\mu \gamma^{\mu\nu}\psi_\nu + {\rm h.c.}
$$
omitted in the $F$--density $\lambda[S_0^3]_F$.

\subsection{Four-dimensional supergravities for all $\,{\cal N}$}

Massless supermultiplets of ${\cal N}$--extended supersymmetry fall in three categories: matter multiplets 
with $|$helicities$|\le1/2$, gauge or Yang-Mills multiplets with $|$helicities$|\le1$
and supergravity multiplets with $|$helicities$|\le2$. The following table indicates, as a function of the number 
${\cal N}$ of supersymmetries:\footnote{The ${\cal N}=7$ theory does not exist: the eighth supersymmetry arises 
automatically and cannot be decoupled.}
\begin{itemize}
\item
The number of (on-shell) helicity states in supergravity, gauge and matter multiplets. 
\item
Supermultiplets with scalar fields. These theories are potentially able, at ground states with nonzero scalar 
expectation values, to offer various patterns of symmetry and supersymmetry breakings
(indicated by $^*$). 
\item
That while Yang-Mills multiplets can gauge all symmetry groups, chirality of the fermion representation can 
only be obtained in the matter (chiral) multiplet of ${\cal N}=1$ supersymmetry.
\end{itemize}
\begin{center}
\begin{tabular}{| c | c | c | c | c | }
\hline
SUSY & Supergravity & $|$Hel.$|$$\le1$ & $|$Hel.$|$ $\le1/2$ & Chirality 
\\
\hline 
$N=1$ & $2_B+2_F$ & $2_B+2_F$ $\,\,\,\,\,$ & $2_B+2_F$ $\,\,^*$ &  {\checkmark} 
\\
$N=2$ & $4_B+4_F$ & $4_B+4_F$ $\,\,^*$& $4_B+4_F$ $\,\,^*$ & - 
\\
$N=3$ & $8_B+8_F$ & $8_B+8_F$ $\,\,^*$& - & - 
\\
$N=4$ & $16_B+16_F\,\,^*$ & $8_B+8_F$ $\,\,^*$& - & - 
\\
$N=5$ & $32_B+32_F\,\,^*$ & - & - & - 
\\
$N=6$ & $64_B+64_F\,\,^*$ & - & - & - 
\\
$N=8$ & $128_B+128_F\,\,^*$ & - & - & - 
\\
\hline
\end{tabular}
\end{center}
Supergravity field theories with ${\cal N}>1$ are much harder to construct. The number of fields of all helicities
increases fast with ${\cal N}$ and off-shell representations do not exist in general. The flexibility in the choice 
of gauge group and matter representation decreases fast with increasing ${\cal N}$. Arbitrary representations 
are allowed with ${\cal N}=1$ only, arbitrary non-chiral representations with ${\cal N}\le2$, and for
${\cal N}\ge3$, only the adjoint representation is admitted. Arbitrary gauge groups are allowed for ${\cal N}\le4$,
while for higher ${\cal N}$, gauged supergravities exclusively depend on the vector fields present in the 
supergravity multiplet. 

The number of vector fields in the supergravity multiplet is ${\cal N}({\cal N}-1)/2$. The choice of possible 
gaugings increases then rapidly with ${\cal N}$, and also taking advantage of electric-magnetic 
duality. These gauged algebras cannot be identified with the compact 
Lie algebras used in the Standard Model or its extensions, but their breaking patterns is a subject of interest 
due to relations with properties found in superstrings. This is an area where fundamental developments of 
supergravity theories is a subject of present researches.

Chirality of the $SU(3)_c\times SU(2)_L \times U(1)_Y$ fermion representation in the Standard Model, 
associated with parity violation by weak interactions, appears to be a fundamental property. This has 
given a particular importance to $N=1$ supergravity coupled to a Yang-Mills multiplet gauging any symmetry 
algebra, and allowing any anomaly-free representation of this gauged symmetry. The most general form of this
theory has been derived in 1982 \cite{CFGVP, ACN}.

\section{ \boldmath{${\cal N}=1$} supergravity--matter couplings} \label{matter}
\setcounter{equation}{0}

The most general interaction of chiral, gauge and supergravity ${\cal N}=1$ multiplets is defined by two 
ingredients. 
Firstly, the choice of a gauge group $G$ and of the representation $R$ of chiral 
supermultiplets. The only constraint would be the absence of chiral anomaly, even if supergravity is not
a quantum field theory. The representation can be chiral and one can then couple the Standard Model to
${\cal N}=1$ supergravity, adding only a sector in which local supersymmetry is spontaneously broken 
(the {\it super-higgs} mechanism \cite{VS, DZ2, Cetal}\footnote{I use the lower case ``higgs" for 
Higgs-Brout-Englert \ldots}) which
should also mediate breaking contributions into the supersymmetric Standard Model (generation of {\it soft breaking terms}).
Secondly, the choice of three gauge-invariant (or gauge-covariant) functions of the scalar 
fields in chiral supermultiplets. The first function, the real K\"ahler potential ${\cal K}$, defines the kinetic 
lagrangian of chiral 
superfields. The holomorphic superpotential $W$ defines the interactions of chiral supermultiplets and the 
holomorphic ${\cal F}$ defines the gauge kinetic (super-Yang-Mills) lagrangian.

These ingredients are known from Poincar\'e global supersymmetry: the most general gauged nonlinear
supersymmetric sigma model is defined in terms of identical ingredients. There is however a subtlety related to
the AdS case. In global supersymmetry in an AdS space-time with cosmological constant $-3M^2$, 
the theory depends on the combination $M{\cal K}
+ W + \ov W$ \cite{AdS}. Then, since supergravity naturally describes AdS {\it and} the limiting Minkowski case,
a similar phenomenon would not be a surprise: one actually finds that the supergravity theory depends 
on\footnote{This combination always used in recent literature
corresponds to $-{\cal G}$ in ref.~\cite{CFGVP}.}
\beq
{\cal G} = {\cal K} + \ln(W\ov W),
\eeq
and of its derivatives (if the superpotential does not vanish). This fact can be loosely traced to the fact that K\"ahler (or $R$) and dilatation symmetries in the superconformal algebra are not compatible (do not
commute) with the AdS superalgebra.

The best procedure to derive the lagrangian is probably\footnote{{\it I.e.} in my opinion.} to start from 
the observation that
all supermultiplets of ${\cal N}=1$ Poincar\'e or AdS supersymmetry are also representations of the ${\cal N}=1$
superconformal symmetry. The method is described in full detail in the book recently published by Dan Freedman 
and Toine Van Proeyen \cite{FVP}. We only give here 
a symbolic explanation and focus on the gravitino and scalar sectors. 
Very schematically, it is as follows:
\begin{itemize}
\item
Consider all supermultiplets denoted as $\Phi^i$ (chiral, helicities $\pm1/2,0,0$) and ${\cal W}_\alpha$ 
(gauge, helicities $\pm1,\pm1/2$) as representations of the superconforma algebra. A Weyl weight and a
$U(1)_R$ charge are then associated with each supermultiplet.
Supergravity fields
$e_\mu^a$ and $\psi_{\alpha\mu}$ are part of superconformal gauge fields.
\item
Add a compensating supermultiplet which is used to gauge fix the unwanted superconformal symmetries.
Here: we symbolically describe old minimal supergravity with a chiral compensating multiplet $S_0$. 
It provides the most general coupling (up to two derivatives and up to some generalizations of minor 
importance) to supergravity \cite{FGKVP}. 
\item
Use tensor calculus methods, as explained in \cite{FVP, KU}, to generate the locally superconformal lagrangian.
\item
Gauge fix superconformal symmetries absent in the Poincar\'e or AdS symmetries 
and eliminate all auxiliary fields. In this step, a gravity frame (Einstein, Jordan, string) is chosen, see below.
\item
Identify the ground state(s) of the theory from the analysis of the scalar potential. It defines the background
geometry (the cosmological constant) and decides if supersymmetry or symmetries in general are 
spontaneously broken.
\end{itemize}
Symbolically, the superconformal lagrangian is represented by
\beq
\label{Lconf1}
{\cal L}=
-{3\over2}\Bigl[ S_0 \ov S_0 \exp\Bigl\{ -{1\over3}{\cal K}(\Phi^i, \ov\Phi_i e^{\cal A}) \Bigr\}\Bigr]_D
+ \Bigl[S_0^3 W(\Phi^i) + {1\over4}{\cal F}(\Phi^i){\cal WW} \Bigr]_F
\eeq
where $[\ldots]_D$ and $[\ldots]_F$ denote the real and chiral invariant densities expressed in
terms of the supermultiplet components and the superconformal gauge fields \cite{FVP, KU}. The Weyl weights 
(scale dimensions) of the supermultiplets are $w= 1$, $0$, $3/2$ for $S_0$, $\Phi^i$, ${\cal W}$ respectively
and the $D$ and $F$ densities apply to supermultiplets with weights 2 and 3: this (with reality and chirality)
dictates the occurences of $S_0$.

\subsection{The scalar sector}

The bosonic part of the lagrangian density (\ref{Lconf1}) also depends on the bosonic
gauge fields of the superconformal algebra, some of them being algebraic (like
the spin connection) or gauge-fixed 
(the dilatation gauge field for instance). The Poincar\'e theory retains the vierbein 
$e_\mu^a$ or metric tensor $g_{\mu\nu}$, the gravitino $\psi_\mu$ and $6_B$ auxiliary fields: 
the gauge field $A_\mu$ of the (gauge-fixed) $U(1)_R$ superconformal symmetry and the complex scalar
$f_0$ in the chiral compensator $S_0$.

After the elimination of all auxiliary fields, a convenient expression for the scalar part of this theory is
\beq
\label{Lconf2}
e^{-1}{\cal L}_{scalar.} = {1\over2}(z_0 \ov z_0{\cal H}) R
-{3\over4}(z_0\ov z_0{\cal H})\left[\partial_\mu{\rm log}(z_0\ov z_0{\cal H})\right]^2
+(z_0\ov z_0{\cal H}){\cal K}^i_j(\partial_\mu z^j)(\partial^\mu \ov z_i)
-V_0
\eeq
with ${\cal H} = \exp[-{\cal K}/3]$ a function of the scalar fields $z^i, \ov z_i$ and with 
\beq
{\cal K}^i_j = {\partial^2\over\partial z^j\partial\ov z_i} {\cal K}.
\eeq
We have kept the complex compensating scalar $z_0$ with scale dimension $w=1$: its value fixes the dilatation and 
$U(1)_R$ gauges. As we can see in the first term, $z_0\ov z_0{\cal H}$ defines the gravity frame,
and the Einstein frame is the gauge condition
\beq
\label{Eins}
{1\over \kappa^2} = z_0\ov z_0 {\cal H} =  z_0\ov z_0 \exp[-{\cal K}/3].
\eeq
In the Einstein frame, 
\beq
\label{Lconf4}
e^{-1}{\cal L}_{scalar.} = {1\over2\kappa^2} R
+ {1\over\kappa^2} \, {\cal K}^i_j(\partial_\mu z^j)(\partial^\mu \ov z_i)
-V_0.
\eeq
The scalar fields $z^i$ are then K\"ahler coordinates: their kinetic metric ${\cal K}^i_j$ derives from the 
{\it K\"ahler potential} ${\cal K}$. Notice that this is only true in the Einstein frame.

The scalar potential is generated by the elimination of auxiliary fields $f^i$ (chiral), $D^A$ (gauge) and $f_0$ 
(in compensator $S_0$). The auxiliary fields 
are:\,\footnote{$T^A_R$: generators of the representation $R$ of chiral multiplets. Possible Fayet-Ilopoulos 
terms are omitted.}
\beq
\label{Lconf3}
\begin{array}{rcl}
f^i &=& -(z_0 \ov z_0{\cal H})^{-1}\, \ov z_0^3 \,({\cal K}^{-1})^i_j \Bigl[  \ov W^j(\ov z_i) + {\cal K}^j\,\ov W(\ov z_i) \Bigr],
\qquad\qquad W_i = {\partial W\over\partial z^i},
\crbig
D^A &=& - [\Re{\cal F}(z^i)]^{-1} (z_0\ov z_0{\cal H}) \, \ov z_i{(T^A_R)^i}_j{\cal K}^j \,,
\crbig
\widetilde f &=& f_0 -{1\over3} z_0\, {\cal K}_i\, f^i  \,\,=\,\,
e^{{\cal K}/3} \, \ov z_0^2 \, \ov W(\ov z_i).
\end{array}
\eeq
And the scalar potential reads
\beq
\label{Lconf6}
V_0 = (z_0 \ov z_0{\cal H}) \, K^i_j \ov f_i f^j + {1\over2} \Re {\cal F} D^A D^A
-3{\cal H} {\tilde f}^*\tilde f.
\eeq
As in global supersymmetry, each auxiliary field $f^i$ of a chiral or $D^A$ of a gauge multiplet produces a positive contribution. A nonzero value at the vacuum state of the potential would spontaneously break supersymmetry.
But, in contrary to the case of global supersymmetry, the supergravity auxiliary field $f^0$ produces, via 
$\widetilde f$, a negative contribution which drives the theory into Anti-de Sitter space. It is then much easier
to create a potential with a ground state breaking supersymmetry, and even at zero cosmological constant.
Notice once again that {\it unbroken supersymmetry} is always associated with either Minkowski or AdS geometry. 

The final form of the scalar potential, before fixing the gravity frame is
\beq
\label{Lconf5}
\begin{array}{rcl}
V_0 &=& \displaystyle
(z_0\ov z_0{\cal H})^2\Bigl\{ e^{\cal K} {\cal K}^i_j [W_i+{\cal K}_iW] [\ov W^j+{\cal K}^j\ov W]
+ {1\over2\Re{\cal F}}\sum_A \, [ \ov z_i{(T^A_R)^i}_j{\cal K}^j]^2
\crbig
&& \hspace{3.5cm} - 3 \,  e^{\cal K} W\ov W
\Bigr\},
\end{array}
\eeq
inserting expressions (\ref{Lconf3}) into the original form (\ref{Lconf6}). In the Einstein frame (\ref{Eins}), 
the prefactor $(z_0\ov z_0{\cal H})^2$ is simply $\kappa^{-4}$. 

But we may as well choose another gravity (Jordan) frame with
\beq
{e^{-2\varphi}\over \kappa^2} = z_0\ov z_0 {\cal H} =  z_0\ov z_0 \exp[-{\cal K}/3].
\eeq
One should understand $\varphi$ as one of the scalar fields in the theory. 
The supergravity lagrangian reads then
\beq
\begin{array}{rcl}
{\cal L} &=& \displaystyle {e^{-2\varphi}\over\kappa^2}e\Bigl[ {1\over2}R
-3(\partial_\mu\varphi)(\partial^\mu\varphi)+{\cal K}^i_j(\partial_\mu z^j)(\partial^\mu \ov z_i) \Bigr]
\crbig
&& -  \displaystyle {e^{-4\varphi}\over\kappa^4}e 
\Bigl[ e^{\cal K} {\cal K}^i_j [W_i+{\cal K}_iW] [\ov W^j+{\cal K}^j\ov W] - 3 e^{\cal K}W\ov W
\crbig
&& \hspace{1.5cm} \displaystyle
+ {1\over2\Re{\cal F}}\sum_A [ \ov z_i{(T^A_R)^i}_j{\cal K}^j]^2\Bigr] + \makebox{gauge and fermion contributions.}
\end{array}
\eeq
Obviously, different frames are related by rescalings of the vierbein field.

In general, one can show that if the scalar potential has a supersymmetric stationary point, 
with values $\langle f^i\rangle = \langle D^A \rangle = 0$ at this point,
it is then stable under small field fluctuations: a supersymmetric vacuum is stable. This does not apply
to non supersymmetric stationary points. 
We will briefly consider some examples with spontaneously broken supersymmetry below. 

\subsection{The gravitino sector}

In the case of pure ${\cal N}=1$ supergravity, we found that a deformation of the Poincar\'e theory leads to a 
negative cosmological constant term associated with an arbitrary energy scale parameter 
$M$. These are eqs.~(\ref{cosmo8}) and (\ref{cosmo6}).

Let us write these gravitino terms as
\beq
e^{-1}{\cal L}_{3/2} = {1\over\kappa^2} \left[ {1\over2} \ov\psi_\mu\gamma^{\mu\nu\rho}\widetilde D_\nu\psi_\rho
+ {1\over2}\, m_{3/2}\, \ov\psi_\mu \gamma^{\mu\nu}\psi_\nu + 3 \, m_{3/2}^2 \right],
\eeq
where $m_{3/2}$ is the quantity appearing in the mass-like term and in the negative or zero 
cosmological constant
\beq
\label{Lis}
\Lambda = -3 m_{3/2}^2.
\eeq
As explained earlier, the relative coefficient $3$ is imposed by supersymmetry and ensures that gravitino
waves have ``light-like" propagation in the AdS geometry. 
In simple Anti-de Sitter supergravity, $m_{3/2}$ is the constant $M$ appearing in the variation 
$\delta\psi_\mu = -{1\over2} M \gamma_\mu\epsilon + \ldots$ 

The supergravity theory coupled to gauge and matter multiplets considered previously and defined by the superconformal expression (\ref{Lconf1}) actually contains a mass-like term for the gravitino with a
field-dependent
\beq
\label{m32is}
m_{3/2} = \kappa^2 |z_0|^3 W = {1\over\kappa} e^{{\cal K}/2} W,
\eeq
if the theory is formulated in the Einstein frame. For a supersymmetric ground state, the expectation value of the 
scalar potential and the induced cosmological constants are then
\beq
\langle V \rangle = -{3\over\kappa^4} \, e^{\cal K} W \ov W 
\qquad\qquad
\Lambda = - \kappa^2 \langle e^{-1}{\cal L}\rangle = \kappa^2 \langle V\rangle = -3\langle |m_{3/2}|^2\rangle,
\eeq
as required. A violation at the vacuum state of these relations would indicate spontaneous supersymmetry 
breaking, and generate a physical mass for the gravitino.

\section{A no-scale model, dilaton supergravity} \label{noscale}

Breaking spontaneously supersymmetry in supergravity is easy. A superpotential is first of all needed.
Here is an example taken in the class of ``no-scale" models \cite{CFKN, LN}.  
Consider a theory describing two chiral supermultiplets with scalar fields $S$ and $T$, 
defined by
\beq
\begin{array}{ll}
{\cal K} = -n\ln(T+\ov T) + \widehat {\cal K}(S,\ov S)  \qquad\qquad &
\makebox{K\"ahler potential,}
\crbig
W = W(S) & \makebox{Superpotential.}
\end{array}
\eeq
The scalar potential reads (using $\kappa=1$ in Einstein frame)
\beq
V = (T+\ov T)^{-n} \, e^{\widehat{\cal K}} \, \left[ \widehat{\cal K}_{S\ov S}^{-1} \,
| W_S +  \widehat{\cal K}_SW |^2 + (n-3) W\ov W \right] .
\eeq
The value $n=3$ is particular: the scalar potential is positive or zero, a solution of
\beq
\label{mineq}
W_S +  \widehat{\cal K}_SW = 0,
\eeq
if it exists, is an absolute minimum and a stable ground state in Minkowski geometry.
This minimum condition fixes in general the value of $\langle S \rangle$
and cancels the auxiliary field $f_S$: supersymmetry is not broken by $S$.
But the value of $\langle T\rangle$ remains arbitrary and the auxiliary field
\beq
f_T = (T+\ov T)^{-1/2} e^{\widehat{\cal K}/2}\, \ov W
\eeq
does not vanish if the superpotential is not zero at the ground state. In this case, supersymmetry is broken
by $T$ and, since the value of $\langle T\rangle$ is not fixed by the potential, the scale of supersymmetry breaking 
is arbitrary and unrelated to any fixed scale of the theory. The gravitino mass
\beq
m_{3/2} = \langle e^{{\cal K}/2} W \rangle = \langle (T+\ov T)^{-3/2} e^{\widehat{\cal K}/2} W \rangle
\eeq
is the order parameter of supersymmetry breaking. Since supersymmetry is broken with zero cosmological 
constant, $m_{3/2}$ is the true gravitino mass.

The K\"ahler potential $-3\ln(T+\ov T)$ commonly appears in compactifications from ten dimensions, 
$T$ being the volume modulus of the compact space. The $S$ field in a superstring context 
could arise from the dilaton scalar (the string coupling field) partner of the metric $g_{\mu\nu}$ and of an antisymmetric tensor $B_{\mu\nu}$. In this case however, the superpotential does not depend on $S$ and eq.~(\ref{mineq})
cannot be solved. The consequence is in general the absence of a ground state.
Non-perturbative corrections are necessary to create the dependence on $S$ and a minkowskian 
ground state.

It is maybe of interest to examine ``dilaton supergravity" more precisely.
In four space-time dimensions, an antisymmetric tensor $B_{\mu\nu}$ with gauge invariance
and (free) wave equation
\beq
\delta B_{\mu\nu} = \partial_\mu \Lambda_\nu - \partial_\nu\Lambda_\mu,
\qquad\qquad
\partial^\mu H_{\mu\nu\rho} = 0, \qquad\qquad H_{\mu\nu\rho} = 3\, \partial_{[\mu}B_{\nu\rho]}
\eeq
describes $3_B$ off-shell (since gauge invariance removes three fields) and $1_B$ on-shell states.
The single massless  on-shell state has of course helicity zero. Combined with a real scalar 
$C$ and a Majorana spinor $\chi_\alpha$, the three fields form an off-shell representation
of supersymmetry without any auxiliary field. Actually, at the level of global ${\cal N}=1$ supersymmetry, 
the variations
\beq
\label{lin1}
\begin{array}{c}
\delta C = i \epsilon\chi -i\ov{\epsilon\chi}, \qquad\qquad\qquad
\delta B_{\mu\nu} = {i\over2\sqrt2}\Bigl(\ov\epsilon[\ov\sigma^\mu, \sigma^\nu]\ov\chi
- \epsilon[\sigma^\mu,\ov\sigma^\nu]\chi \Bigr) ,
\crbig
\delta\chi_\alpha = -i(\sigma^\mu\ov\epsilon)_\alpha 
\Bigl({1\over\sqrt2}\epsilon_{\mu\nu\rho\sigma}H^{\nu\rho\sigma} - i \partial_\mu C \Bigr)
\end{array}
\eeq
close the supersymmetry algebra without using field equations. This {\it linear} multiplet can be coupled to
supergravity and is also a representation of the superconformal algebra with Weyl weight $w=2$ for $C$.
The variations (\ref{lin1}) indicate that a constant $C$ does not break supersymmetry and that the spinor 
$\chi$ cannot be a Goldstino spinor. Hence a linear multiplet is not a source for supersymmetry breaking
and it does not contribute to the scalar potential.

A duality transformation can always, in principle, transform the antisymmetric tensor with gauge symmetry into
a real scalar $\tau$ with shift symmetry. Schematically,
\beq
H_{\mu\nu\rho} = 3\, \partial_{[\mu}B_{\nu\rho]} \qquad\longleftrightarrow \qquad
\epsilon_{\mu\nu\rho\sigma} \partial^\sigma \tau
\eeq
with symmetry $\delta\tau =$ constant. In other words, the dual chiral theory has a K\"ahler potential
${\cal K}(S+\ov S)$, $\tau = \Im S$. 

On one side of the duality, the linear $L$ does not have an auxiliary field and cannot break supersymmetry. 
On the other side, the chiral dual $S$ has an auxiliary $f_S$ in principle able to induce supersymmetry breaking. 
In global supersymmetry, what happens is that either $f_S$ is identically zero or, if other chiral multiplets are present,
$f_S$ is a linear combination of the other auxiliary fields $f^i$. In any case, $f_S$ does not provide an {\it independent}
source of supersymmetry breaking. Hence, models with supersymmetry breaking induced exclusively
by the chiral $S$ dual to the string dilaton and the antisymmetric tensor $B_{\mu\nu}$ do not exist.

In supergravity with a linear multiplet, the situation is different. If the superpotential is constant (this is 
meaningless in global supersymmetry), there is always a scalar potential generated by the supergravity 
auxiliary field $f_0$ (which however cannot break supersymmetry).
In the dual version with the chiral $S$, the auxiliary field $f_S$ is proportional to $f_0$, but $f_S$ is now
in principle able to break supersymmetry, since $\ov f_S \sim {\cal K}_T W$ is not zero in general. 
What happens now if that the potential is in general unstable if $W\ne0$: this is the runaway behaviour 
naturally expected from the dilaton, which in turn raises the problem of its stabilisation. Notice that supergravity with
the K\"ahler potential $-3\ln (T+\ov T)$, which has identically zero potential and broken supersymmetry if $W\ne0$,
{\it cannot} be transformed into a linear superfield: the supersymmetric duality transformation between
$S$ and $L$ does not exist for precisely this K\"ahler potential. If other chiral multiplets are present, 
$f_S$ is a linear combination of the chiral auxiliary fields $f^i$ and of the supergravity $f_0$. Again, $f_S$
is not an independent source of breaking. But stability remains a non simple issue.

The simplest Calabi-Yau compactifications of heterotic superstrings with ${\cal N}=1$ four-dimensional 
supersymmetry provide a concrete realization of the mechanisms described in this section.
Retaining the overall (complex) volume modulus $T$ in a chiral multiplet and the dilaton--antisymmetric tensor supermultiplet, we have two dual descriptions, 
$$
L \quad \makebox{(linear) \quad and} \quad T \quad\qquad\Longleftrightarrow 
\quad\qquad S \quad\makebox{(chiral) \quad and}\quad T.
$$
In the chiral version, at lowest order of string perturbation theory, the K\"ahler potential defining the effective
supergravity is of no-scale type \cite{DIN1, W, DRSW, DIN2},
\beq
{\cal K} = -\ln (S+\ov S) - 3\ln(T+\ov T).
\eeq
There are two primary sources for a superpotential \cite{DIN1, DRSW}. Firstly, at the perturbative level, 
ten-dimensional sixteen-supercharge supergravity has a gauge-invariant three-form $H_{MNP}$. It generates
$H_{\mu\nu\rho} = 3\,\partial_{[\mu}B_{\nu\rho]}$ in the dilaton sector and an order parameter 
$H_{ijk}$, leading to a constant superpotential $W=\langle H\rangle$, since the Calabi-Yau space has a holomorphic three-form. 
At this stage, the minimum equation (\ref{mineq}) cannot be solved and the potentiel 
\beq
V = (T+\ov T)^{-3}(S+\ov S)W\ov W
\eeq
is unstable (``run-away behaviour"), as expected from dilaton supergravity. 
The second source of superpotential is nonperturbative gaugino condensation in a hidden 
gauge sector, and it is described to a good approximation by the addition to the constant $\langle H\rangle$ 
of a term of the form $a e^{bS}$:
\beq
W(S) = \langle H\rangle + a e^{bS}.
\eeq
The equation $f_S=0$ can now be solved and supersymmetry breaks in the $T$ sector, with Minkowski
geometry and order parameter (\ref{m32is}) controlled by the arbitrary value of $T+\ov T$.

It should however be observed that heterotic string perturbation theory is organized in powers of the linear 
supermultiplet $L$ \cite{CFV}, with its scalar $C$ directly related to the string dilaton, and not as an expansion in $S$.
Since the superpotential cannot depend on $L$, the description of gaugino condensation uses then a different 
effective lagrangian, with almost identical phenomenology as long as supersymmetry breaking and scales are concerned \cite{BDQQ}. In any case, the $S\sim L$ duality is a useful tool in the effective description of the 
universal string dilaton sector.

\section{Final words}

After (almost) forty years, supergravity has certainly found its way into the toolbox of theoretical physicists. 
Its development is far from complete and gauge-gravity dualities, in particular, have recently suggested new directions
and research projects. In the context of superstring compactifications, finding methods to classify supergravity 
gaugings, and the corresponding symmetry and supersymmetry breaking patterns would allow a better control
of flux compactifications, with supergravity providing in this case the ``bottom-up" approach to select candidate
fluxes from specific low-energy properties. In the unification programme, supergravity cannot claim to be the
fundamental theory. But it is certainly on the ``supersymmetric path"
to quantum gravity. And there are open questions concerning the quantum status of the maximal ${\cal N}=8$
supergravity. Supersymmetric theories have exceptional ultraviolet properties. The maximal (${\cal N}=4$)
super-Yang-Mills theory is known to be finite. Brilliant works have shown that divergences plausible in ${\cal N}=8$
arise in perturbation theory at higher orders than expected. This suggests that the ${\cal N}=8$ theory could maybe
display ingredients of a consistent quantum gravity, in a much simpler theoretical framework
than superstrings.

\section*{Acknowledgements}

I wish to thank the organizers of Discrete 2014, and especially Nick Mavromatos, for the invitation to present in 
their conference a subject which fascinates me since many years. This work has been in part supported by the
Swiss National Science Foundation,

\end{document}